\documentclass[10pt,final,journal]{IEEEtran}
\usepackage[switch]{lineno} 
\usepackage{tabularx} 
\usepackage{booktabs}
\usepackage{multirow}
\usepackage{makecell}

\usepackage{stmaryrd}
\usepackage{graphicx}
\usepackage{amsthm} 
\usepackage{amsmath} 
\usepackage{bm}
\usepackage{graphicx}
\usepackage{epstopdf}
\usepackage{color}
\usepackage{float}
\usepackage[colorlinks=false,linkcolor=black,urlcolor=black,bookmarksopen=true, hidelinks]{hyperref}
\usepackage{bookmark}
\usepackage[flushleft]{threeparttable}
\usepackage{stackengine}
\usepackage{scalerel}
\usepackage{array}

\usepackage{xcolor,soul}

\usepackage{cite}

\ifCLASSINFOpdf
\else
\fi
\usepackage{amsfonts,amssymb}

\usepackage{amsmath}
\usepackage{algorithm}
\usepackage{algpseudocode}
\ifCLASSOPTIONcompsoc
 \usepackage[caption=false,font=normalsize,labelfont=sf,textfont=sf]{subfig}
\else
 \usepackage[caption=false,font=footnotesize]{subfig}
\fi
\begin{document}
%

\title{Closed-form Two-way TOA Localization and Synchronization for User Devices with Motion and Clock Drift}

\author{Sihao~Zhao, 
		Ningyan~Guo,
		Xiao-Ping~Zhang, \textit{Fellow, IEEE},
        Xiaowei~Cui,
        and~Mingquan~Lu
\thanks{This work was supported in part by the Natural Sciences and Engineering Research Council of Canada (NSERC), Grant No. RGPIN-2020-04661. \textit{(Corresponding author: Xiao-Ping Zhang)}}
\thanks{S. Zhao and X.-P. Zhang are with the Department of Electrical, Computer and Biomedical Engineering, Ryerson University, Toronto, ON M5B 2K3, Canada (e-mail: sihao.zhao@ryerson.ca; xzhang@ryerson.ca).}
\thanks{N. Guo, X. Cui and M. Lu are with the Department of Electronic Engineering,
	Tsinghua University, Beijing 100084, China (e-mail: guoningyan@tsinghua.edu.cn; cxw2005@tsinghua.edu.cn; lumq@tsinghua.edu.cn).}
}

\markboth{}%
{Shell \MakeLowercase{\textit{et al.}}: Bare Demo of IEEEtran.cls for IEEE Journals}
%




\maketitle

\begin{abstract}
A two-way time-of-arrival (TOA) system is composed of anchor nodes (ANs) and user devices (UDs). Two-way TOA measurements between AN-UD pairs are obtained via round-trip communications to achieve localization and synchronization (LAS) for a UD. Existing LAS method for a moving UD with clock drift adopts an iterative algorithm, which requires accurate initialization and has high computational complexity. In this paper, we propose a new closed-form two-way TOA LAS approach, namely CFTWLAS, which does not require initialization, has low complexity and empirically achieves optimal LAS accuracy. We first linearize the LAS problem by squaring and differencing the two-way TOA equations. We employ two auxiliary variables to simplify the problem to finding the analytical solution of quadratic equations. Due to the measurement noise, we can only obtain a raw LAS estimation from the solution of the auxiliary variables. Then, a weighted least squares step is applied to further refine the raw estimation. We analyze the theoretical error of the new CFTWLAS and show that it empirically reaches the Cram\'er-Rao lower bound (CRLB) with sufficient ANs under the condition of proper geometry and small noise. Numerical results in a 3D scenario verify the theoretical analysis that the estimation accuracy of the new CFTWLAS method reaches CRLB in the presented experiments when the number of ANs is large, the geometry is appropriate, and the noise is small.
Unlike the iterative method whose complexity increases with the iteration count, the new CFTWLAS has constant low complexity.

\end{abstract}

\begin{IEEEkeywords}
two-way time-of-arrival (TOA), closed-form, localization and synchronization (LAS), motion, clock drift.
\end{IEEEkeywords}


%
\IEEEpeerreviewmaketitle

\section{Introduction}\label{Introduction}
%
%
%
%
\IEEEPARstart{L}{ocalization} and synchronization (LAS) techniques provide position and timing information, which is significant to a variety of real-world applications such as Internet of Vehicles, Internet of Things (IoT), emergency rescue and surveillance reconnaissance \cite{kuutti2018survey,beard2002coordinated}. Among the measurements such as time-of-arrival (TOA), angle-of-arrival (AOA) and received signal strength (RSS) \cite{shao2014efficient,hu2017robust,luo2019novel,coluccia2019hybrid,wang2012novel,an2020distributed}, TOAs between the anchor nodes (ANs) and a user device (UD) are widely adopted in LAS and different types of methods have been developed to solve the LAS problem \cite{chan1994simple,wang2019convex,zhao2020closed,vaghefi2015cooperative,zhao2021opti,shi2019blas,chepuri2012joint,zhao2021parn,rui2014elliptic,zhao2021optimal}. A typical example of such a scheme is the widely used Global Positioning System (GPS) \cite{misra2006global,borre2007software,zhao2014kalman}. 

Two TOA measurements or two-way TOA can be obtained if there is round-trip communication between a pair of AN and UD. Two-way TOA based LAS techniques have been studied and implemented extensively \cite{zheng2010joint,denis2006joint,cano2019kalman,ouguz2013tw,gholami2016tw,gao2016robust,zhao2020optimaltwo,bialer2016two,tomic2018exact}. However, these studies assume that there is no UD clock drift and/or the UD is stationary. Ignoring the UD clock drift will result in large LAS error in applications using low cost clock sources. And the assumption of a stationary UD limits the applications of the two-way TOA LAS techniques in moving cases such as personnel/asset tracking, autonomous vehicle navigation and wearable IoT device localization.

Recent research work, taking both the UD velocity and clock drift into account, formulates the LAS problem as a maximum likelihood (ML) estimator and develops a Gauss-Newton iterative method to solve it \cite{zhao2020optimaltwo}. However, the iterative method strongly depends on a good initial guess to achieve the optimal estimation, and has high computational complexity due to iterations. Another work relaxes the ML estimator to a semidefinite programming problem \cite{zhao2021sdptwtoa}. However, its solution is suboptimal and it also suffers from high complexity.

In this paper, to solve the LAS problem in a two-way TOA scenario with UD motion and clock drift, we propose a new closed-form LAS method utilizing two-way TOA measurements, namely CFTWLAS. Inspired by the idea of converting the localization problem in the sequential broadcast one-way TOA case to solving a quadratic equation set in \cite{guo2021new}, we devise two auxiliary variables to simplify the LAS problem in this two-way TOA scenario into finding the solution of a quadratic equation set. We first square the two-way TOA measurement equations to obtain linear relations. Then we construct two transformation matrices to connect the linear relations with the two auxiliary variables and form two quadratic equations. After obtaining the roots of the quadratic equations analytically, we obtain the raw estimate of the LAS parameters. We further apply a refinement step based on weighted least squares (WLS) to obtain the optimized LAS result. We show that under small noise and far field conditions, the estimation error empirically reaches Cram\'er-Rao lower bound (CRLB). We conduct numerical simulation in a 2D scene. Results verify the theoretical analysis that with sufficient ANs, the CFTWLAS is empirically optimal under small noise and far field conditions in the presented experiments. Compared with the conventional iterative method, the new CFTWLAS does not require initialization to ensure a correct solution. In addition, the CFTWLAS has fixed low complexity, in contrast to the increasing complexity of the iterative method when the number of iteration grows.

\section{Problem Formulation} \label{problem}
There are $M$ anchor nodes (ANs) with known $N$-dimensional ($N=$2 or 3) coordinate is denoted by $\boldsymbol{q}_i$, $i=1,\cdots,M$. All the ANs' clocks are synchronized using methods such as multiple timestamp exchanges between ANs \cite{shi2019blas}.

\begin{figure}
	\centering
	\includegraphics[width=0.90\linewidth]{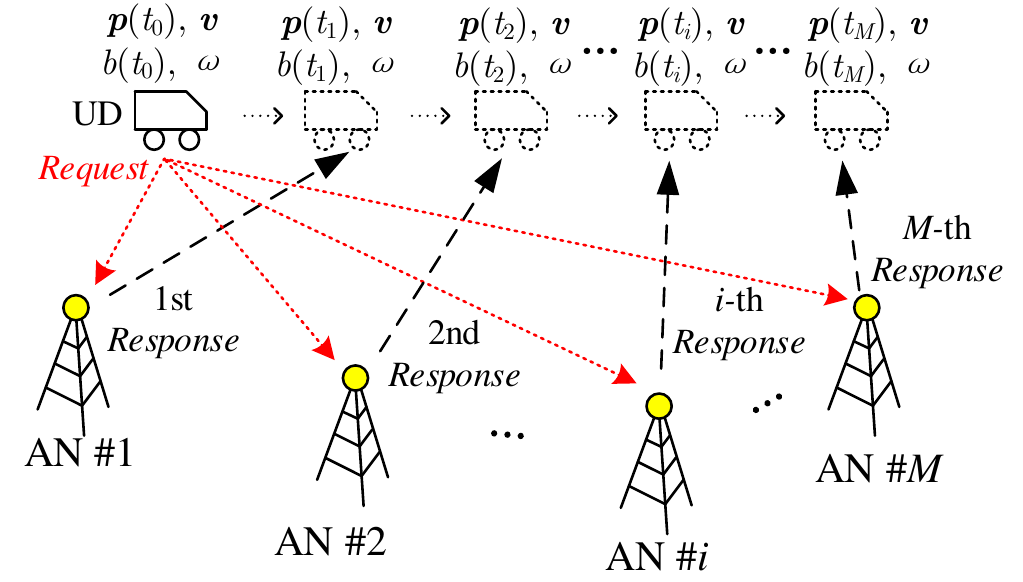}
	\vspace{-0.3cm}
	\caption{Two-way TOA localization system. The moving UD transmits the request signal, and all ANs receive to form $M$ request-TOA measurements. Then, the ANs transmit response signals sequentially. The UD receives to form $M$ sequential response-TOA measurements.
	}
	\label{fig:systemfig}
	\vspace{-0.3cm}
\end{figure}

There can be multiple user devices (UDs) in this two-way TOA system. Without loss of generality, we take one UD as an example. We denote its unknown parameters including position, velocity, clock offset and clock drift at the time instant $t_0$ by $\boldsymbol{p}$, $\boldsymbol{v}$, $b$, and $\omega$, respectively. As shown in Fig. \ref{fig:systemfig}, the UD communicates with the ANs while it moves. Specifically, at time instant $t_0$, the UD first transmits the request signal, and the $M$ ANs receive this signal. Then, ANs transmit the response signals sequentially, which are received by the UD at $t_1$, $\cdots$, $t_M$, respectively. During this short period of communication, the UD's velocity and clock drift are considered constant.


%

When AN \#$i$ ($i=1,\cdots,M$) receives the request signal from the UD, the request-TOA measurement, denoted by $\hat{\rho}_i$, is modelled as
\begin{align} \label{eq:rhoANi}
	\hat{\rho}_i = \rho_i+\varepsilon_{i}=
	\left\Vert\boldsymbol{q}_i-\boldsymbol{p}\right\Vert -b+ \varepsilon_{i} \text{, } i=1,\cdots,M \text{,}
\end{align}
where all the time-related terms are multiplied by the signal propagation speed and have the unit of meter, $\rho_i$ is the noise free version of the request-TOA, $\varepsilon_{i}$ is the measurement noise for AN \#$i$, following independent zero mean Gaussian distribution with a variance of $\sigma_{i}^2$, i.e., $\varepsilon_{i} \sim \mathcal{N}(0,\sigma_{i}^2)$.

The response-TOA measurement, denoted by $\hat{\tau}_i$, is obtained when the UD receives the response signal from AN \#$i$. It is determined by the true distance between the UD and AN \#$i$ and the clock offset at the instant of reception plus measurement noise. We write $\tau_i$ as
\begin{align} \label{eq:tauANi}
	\hat{\tau}_i =\tau_i+\eta_i= \left\Vert\boldsymbol{q}_i-\boldsymbol{p}-\boldsymbol{v}\Delta t_i\right\Vert+ b+\omega\Delta t_i + \eta_i \text{, }
\end{align}
where $\tau_i$ is the noise free version of the response-TOA, $\Delta t_i$ is the interval between the transmission of the request signal and the reception of the response signal from AN \#$i$, i.e., $\Delta t_i=t_i-t_0$, $\eta_i$ is the measurement noise for the UD, following a zero-mean Gaussian distribution with a variance of $\sigma^2$, i.e., $\eta_i \sim \mathcal{N}(0,\sigma^2)$, and $i=1,\cdots,M$.

The LAS problem for a moving UD with clock drift is to estimate the coordinate $\boldsymbol{p}$ and clock offset $b$ at the instant $t_{0}$, given the two-way TOA measurements in (\ref{eq:rhoANi}) and (\ref{eq:tauANi}). We will develop a new closed-form LAS method for this problem, namely CFTWLAS, in the next section.

\section{Closed-form Two-way Localization and Synchronization Method (CFTWLAS)}

\subsection{Step 1: Linearization}
The relation between the TOA measurements and the UD parameters including position, velocity, clock offset and drift given by (\ref{eq:rhoANi}) and (\ref{eq:tauANi}) is nonlinear. We first consider the noise free version of $\rho_i$ and $\tau_i$. To obtain a linear relation, we re-organize and take a square and have
\begin{subequations}
\begin{align}
\left({\rho}_i+b\right)^2&=\left\Vert\boldsymbol{p}-\boldsymbol{q}_i \right\Vert^2,	\label{eq:rhosquare}\\
\left({\tau}_i-b-\omega \Delta t_i\right)^2&=\left\Vert\boldsymbol{p}+\boldsymbol{v} \Delta t_i-\boldsymbol{q}_i \right\Vert^2. 	\label{eq:tausquare}
\end{align}
\end{subequations}

After re-organizing, they become
\begin{subequations}
\label{eq:rhotausquare}
\begin{align}
	&\rho_i^2+2\rho_i b+\left(b^2-\Vert\boldsymbol{p}\Vert^2\right)+2\boldsymbol{q}_i^T\boldsymbol{p}=\Vert\boldsymbol{q}_i\Vert^2, \label{eq:rhosquare1}\\
	&2\boldsymbol{q}_i^T\boldsymbol{p}+2\Delta t_i\boldsymbol{q}_i^T\boldsymbol{v}-2{\tau}_i b-2\Delta t_i{\tau}_i\omega+\left(b^2-\Vert\boldsymbol{p}\Vert^2\right)\\
	&=\Vert\boldsymbol{q}_i\Vert^2-{\tau}_i^2-\Delta t_i^2\left(\omega^2-\Vert\boldsymbol{v}\Vert^2\right)-2\Delta t_i\left(b\omega-\boldsymbol{p}^T\boldsymbol{v}\right),\nonumber
\end{align}
\end{subequations}


To remove the term $b^2-\Vert\boldsymbol{p}\Vert^2$, without loss of generality, we substitute $i=1$ into (\ref{eq:rhotausquare}), then subtract it from (\ref{eq:rhotausquare}) that has other values of $i$, and obtain
\begin{subequations}
\label{eq:diffeq}
\begin{align}
	&(\rho_i-\rho_1)b+(\boldsymbol{q}_i^T-\boldsymbol{q}_1^T)\boldsymbol{p}\nonumber\\
	=&\frac{1}{2}\Vert\boldsymbol{q}_i\Vert^2-\frac{1}{2}\Vert\boldsymbol{q}_1\Vert^2-\frac{1}{2}(\rho_i^2-\rho_1^2),\\
	&\left(\boldsymbol{q}_i^T-\boldsymbol{q}_1^T\right)\boldsymbol{p}+\left(\Delta t_i\boldsymbol{q}_i^T-\Delta t_1\boldsymbol{q}_1^T\right)\boldsymbol{v}+\left({\tau}_1-{\tau}_i\right)b \nonumber\\
	+&\left(\Delta t_1{\tau}_1-\Delta t_i{\tau}_i\right)\omega\nonumber\\
	=&\frac{1}{2}\Vert\boldsymbol{q}_i\Vert^2-\frac{1}{2}\Vert\boldsymbol{q}_1\Vert^2-\frac{1}{2}\left({\tau}_i^2-{\tau}_1^2\right)\nonumber\\
	+&\frac{1}{2}\left(\Delta t_1^2-\Delta t_i^2\right)\left(\omega^2-\Vert\boldsymbol{v}\Vert^2\right)+ \left(\Delta t_1-\Delta t_i\right)\left(b\omega-\boldsymbol{p}^T\boldsymbol{v}\right),\label{eq:difftausquare} \nonumber\\
	&i=2,\cdots,M.
\end{align}
\end{subequations}

We denote the unknown parameters by $\boldsymbol{\theta}=[\boldsymbol{p}^T,\boldsymbol{v}^T,b,\omega]^T$, and employ two auxiliary variables as
\begin{align}
	\lambda_1=\omega^2-\Vert\boldsymbol{v}\Vert^2,\; \text{and }
	\lambda_2=b \omega-\boldsymbol{p}^T\boldsymbol{v}.
\end{align}
We then rewrite (\ref{eq:diffeq}) into the collective form as
\begin{align}\label{eq:collective1}
	\bm{A}\boldsymbol{\theta}=\boldsymbol{y}+\bm{G}\left[\lambda_1,\lambda_2\right]^T,
\end{align}
in which
$
	\bm{A}=[\bm{A}_{\rho}^T,\;\bm{A}_{\tau}^T]^T, \; \boldsymbol{y}=[\boldsymbol{y}_{\rho}^T,\boldsymbol{y}_{\tau}^T]^T, \bm{G}=[\bm{G}_{\rho}^T,\bm{G}_{\tau}^T]^T, \;
	[\bm{A}_{\rho}]_{i,:}=[\boldsymbol{q}_{i+1}^T-\boldsymbol{q}_{1}^T,\bm{0}_N^T,\rho_{i+1}-\rho_1,0],\;
	[\bm{A}_{\tau}]_{i,:}=
	[\boldsymbol{q}_{i+1}^T-\boldsymbol{q}_1^T, \Delta t_{i+1}\boldsymbol{q}_{i+1}^T-\Delta t_1\boldsymbol{q}_1^T, {\tau}_1-{\tau}_{i+1},\Delta t_1{\tau}_1-\Delta t_{i+1}\tau_{i+1}],\;
	[\boldsymbol{y}_\rho]_i=
	\left(\Vert\boldsymbol{q}_{i+1}\Vert^2-\Vert\boldsymbol{q}_1\Vert^2-\left({\rho}_{i+1}^2-{\rho}_1^2\right)\right)/2,
	\;
	[\boldsymbol{y}_\tau]_i=
	\left(\Vert\boldsymbol{q}_{i+1}\Vert^2-\Vert\boldsymbol{q}_1\Vert^2-\left({\tau}_{i+1}^2-{\tau}_1^2\right)\right)/2,\;
	[\bm{G}_\rho]_{i,:}=[0,0],\;
	[\bm{G}_\tau]_{i,:}=\left[
	\begin{matrix}
		\frac{1}{2}(\Delta t_1^2-\Delta t_{i+1}^2), & \Delta t_1-\Delta t_{i+1}
	\end{matrix}
	\right],
$
where $[\cdot]_{i,:}$ represents the $i$-th row of a matrix, and $[\cdot]_{i}$ is the $i$-th element of a vector, and $i=1,\cdots,M-1$.

At this stage, we have obtained the linear relation of the unknown parameter $\boldsymbol{\theta}$ and the two auxiliary variables $\lambda_1$ and $\lambda_2$ as given by (\ref{eq:collective1}).
To determine $\boldsymbol{\theta}$, we will then find the solution of $\lambda_1$ and $\lambda_2$ in the next subsection.

\subsection{Step 2: Raw Estimation}
We obtain $\boldsymbol{\theta}$ with respect to $\bm{\lambda}=\left[\lambda_1,\lambda_2\right]^T$ by applying a least squares (LS) method to (\ref{eq:collective1}), as given by
\begin{align}\label{eq:LSestimate}
{\boldsymbol{\theta}}=\left(\bm{A}^T\bm{A}\right)^{-1}\bm{A}^T\left(\boldsymbol{y}+\bm{G}\boldsymbol{\lambda}\right)
=\boldsymbol{g}+\bm{U}\boldsymbol{\lambda},
\end{align}
where
$
	\boldsymbol{g}=\left(\bm{A}^T\bm{A}\right)^{-1}\bm{A}^T\boldsymbol{y},\;
	\bm{U}=\left(\bm{A}^T\bm{A}\right)^{-1}\bm{A}^T\bm{G}.
$

Note that although \eqref{eq:LSestimate} has the same form, the variables $\bm{A}$, $\boldsymbol{y}$, $\bm{G}$, $\boldsymbol{g}$ and $\bm{U}$ are different from those in \cite{guo2021new} due to the two-way TOA scenario, which brings doubled number of measurements and leads to more complex processing of the measurements.

To obtain the solution of \eqref{eq:LSestimate}, $\bm{A}$ must be full column rank for inverse operation of $\bm{A}^T\bm{A}$. Thus, we need at least $N+2$ available ANs and a proper AN geometry to ensure its full rank. An extreme case is when the UD is at the center of a 2D square area with 4 ANs placed at the 4 corners. Such cases will cause rank deficiency in $\bm{A}$ and must be avoided.

To find the solution of the auxiliary variables $\lambda_1$ and $\lambda_2$, we first construct the following two matrices as
\begin{align}
	\bm{H}_1&=\mathrm{diag}\left([\boldsymbol{0}_N^T,-\boldsymbol{1}_N^T,0,1]\right),\label{eq:H1}\\
	\bm{H}_2&=\left[
	\begin{matrix}
		\bm{O}_{N\times N} &-\bm{I}_{N}&\bm{O}_{N\times 2}\\
		-\bm{I}_{N} & \bm{O}_{N\times N} & \bm{O}_{N\times 2}\\
		\bm{O}_{2\times N}  & \bm{O}_{2\times N} &
		\begin{matrix}
			0&1\\
			1&0
		\end{matrix}
	\end{matrix}
	\right],
	\label{eq:H2}
\end{align}
where $\mathrm{diag}(\cdot)$ is diagonal matrix with the elements of the vector inside as the diagonal entries, $\boldsymbol{0}_N$ and $\boldsymbol{1}_N$ are vectors with all zeros and ones, respectively, $\bm{O}_{N\times N}$ is $N\times N$ zero matrix, and $\bm{I}_{N}$ is $N\times N$ identity matrix.

Based on \eqref{eq:LSestimate}, (\ref{eq:H1}) and (\ref{eq:H2}) we have
\begin{align}
	&\left(\boldsymbol{g}+\bm{U}\boldsymbol{\lambda}\right)^T\bm{H}_1\left(\boldsymbol{g}+\bm{U}\boldsymbol{\lambda}\right)
	=\omega^2-\Vert\boldsymbol{v}\Vert^2=\lambda_1,\label{eq:H1a}\\
	&\left(\boldsymbol{g}+\bm{U}\boldsymbol{\lambda}\right)^T\bm{H}_2\left(\boldsymbol{g}+\bm{U}\boldsymbol{\lambda}\right)
	=2\left(b\omega-\boldsymbol{p}^T\boldsymbol{v}\right)=2\lambda_2. \label{eq:H2a}
\end{align}

After re-organizing (\ref{eq:H1a}) and (\ref{eq:H2a}), we obtain two quadratic equations with respect to $\lambda_1$ and $\lambda_2$ as
\begin{subequations}
\label{eq:equationset}
\begin{align} 
	a_1 \lambda_1^2+b_1 \lambda_1\lambda_2 +c_1 \lambda_2^2 + d_1 \lambda_1+e_1 \lambda_2+f_1&=0 \text{,} \label{eq:quadratic1}\\
	a_2 \lambda_1^2+b_2 \lambda_1\lambda_2 +c_2 \lambda_2^2 + d_2 \lambda_1+e_2 \lambda_2+f_2&=0 \text{,}\label{eq:quadratic2}
\end{align}
\end{subequations}
where
$
	a_1=[\bm{U}]_{:,1}^T\bm{H}_1[\bm{U}]_{:,1}, \; b_1=2[\bm{U}]_{:,1}^T\bm{H}_1[\bm{U}]_{:,2}, \;
	c_1=[\bm{U}]_{:,2}^T\bm{H}_1[\bm{U}]_{:,2}, \;
	d_1=2[\bm{U}]_{:,1}\bm{H}_1\boldsymbol{g}-1,\;
	e_1=2[\bm{U}]_{:,2}\bm{H}_1 \boldsymbol{g},\;
	f_1=\boldsymbol{g}^T\bm{H}_1\boldsymbol{g},\;
	a_2=[\bm{U}]_{:,1}^T\bm{H}_2[\bm{U}]_{:,1},\;
	b_2=2[\bm{U}]_{:,1}^T\bm{H}_2[\bm{U}]_{:,2},\;
	c_2=[\bm{U}]_{:,2}^T\bm{H}_2[\bm{U}]_{:,2},\;
	d_2=2[\bm{U}]_{:,1}\bm{H}_2\boldsymbol{g},\;
	e_2=2[\bm{U}]_{:,2}\bm{H}_2\boldsymbol{g}-2,\;
	f_2=\boldsymbol{g}^T\bm{H}_2\boldsymbol{g},
$
in which $[\cdot]_{:,j}$ represents the $j$-th column of a matrix.

The equation set \eqref{eq:equationset} can be solved analytically following \cite{zhao2020closed}. Then, the parameter ${\boldsymbol{\theta}}$ can be obtained by (\ref{eq:LSestimate}).

Note that until this stage, we consider the solution without measurement noise. In practice, we only have the noisy version of the request and response TOA measurements. Therefore, in the above steps, the matrix $\bm{A}$ and vector $\boldsymbol{y}$ are approximated by the noisy TOA measurements $\hat{\rho}_i$ and $\hat{\tau}_i$. As a result, the parameter obtained from (\ref{eq:LSestimate}) is a raw estimate, denoted by $\tilde{\boldsymbol{\theta}}=[\tilde{\boldsymbol{p}}^T,\tilde{\boldsymbol{v}}^T,\tilde{b},\tilde{\omega}]^T$.

There may be multiple roots of $\lambda_1$ and $\lambda_2$ from the equation set \eqref{eq:equationset} and thus there may be more than one estimates of $\tilde{\boldsymbol{\theta}}$. We select the one that satisfies
$
\min_{\tilde{\boldsymbol{\theta}}} [\boldsymbol{r}_{\rho}^T,\boldsymbol{r}_{\tau}^T]\bm{W}[\boldsymbol{r}_{\rho}^T,\boldsymbol{r}_{\tau}^T]^T
$,
where $\bm{W}=\mathrm{diag}\left(\left[1/\sigma_1^2,\cdots,1/\sigma_M^2, \boldsymbol{1}_M^T1/\sigma^2\right]\right),
$
and
$[\boldsymbol{r}_{\rho}]_{i}=\hat{\rho}_i-\left\Vert {\boldsymbol{q}}_i-\tilde{\boldsymbol{p}}\right\Vert +\tilde{b},
[\boldsymbol{r}_{\tau}]_{i}=\hat{\tau}_i-\left\Vert \boldsymbol{q}_i-\tilde{\boldsymbol{p}}-\tilde{\boldsymbol{v}}\Delta t_i\right\Vert -\tilde{b}-\tilde{\omega}\Delta .
$


\subsection{Step 3: WLS Refinement}
The raw estimate $\tilde{\boldsymbol{\theta}}$ is not optimal when there are measurement noises. When the raw estimation $\tilde{\boldsymbol{\theta}}$ is not far from the true parameter $\boldsymbol{\theta}$ under the small noise and far field conditions, we can use the first order term of the Taylor series to express the collective form of TOA measurements $\boldsymbol{\gamma}$ as
\begin{align}\label{eq:ftheta}
	\boldsymbol{\gamma}= \mathit{h}(\tilde{\boldsymbol{\theta}}) + \left(\frac{\partial \mathit{h}(\boldsymbol{\theta})}{\partial \boldsymbol{\theta}}|_{\boldsymbol{\theta}=\tilde{\boldsymbol{\theta}}}\right)\left(\boldsymbol{\theta}-\tilde{\boldsymbol{\theta}}\right)+[\boldsymbol{\varepsilon}^T,\boldsymbol{\eta}^T]^T,
\end{align}
where $\boldsymbol{\gamma}=[\hat{\rho}_1,\cdots,\hat{\rho}_M,\hat{\tau}_1,\cdots,\hat{\tau}_M]^T$, $h(\boldsymbol{\theta})=[h_{\rho}^T(\boldsymbol{\theta}), h_{\tau}^T(\boldsymbol{\theta})]^T$ is a function of $\boldsymbol{\theta}$, 
$
\left[h_{\rho}(\boldsymbol{\theta})\right]_i=\left\Vert\boldsymbol{q}_{i}-\boldsymbol{p}\right\Vert-b,\;
\left[h_{\tau}(\boldsymbol{\theta})\right]_i=\left\Vert\boldsymbol{q}_{i}-\boldsymbol{p}-\boldsymbol{v}\Delta t_{i}\right\Vert+b+\omega \Delta t_{i},\;
\text{.}
$, $\boldsymbol{\varepsilon}$ and $\boldsymbol{\eta}$ are the collective form of $\varepsilon_i$ and $\eta_i$.

We denote $\tilde{\bm{J}}=\frac{\partial \mathit{h}(\boldsymbol{\theta})}{\partial \boldsymbol{\theta}}|_{\boldsymbol{\theta}=\tilde{\boldsymbol{\theta}}}=[\tilde{\bm{J}}_{\rho}^T,\tilde{\bm{J}}_{\rho}^T]^T$, where
$[\tilde{\bm{J}}_{\rho}]_{i,:}=\left[\frac{\partial \mathit{h}_{\rho}(\boldsymbol{\theta})}{\partial \boldsymbol{\theta}}|_{\boldsymbol{\theta}=\tilde{\boldsymbol{\theta}}}\right]_{i,:} =\left[-\tilde{\boldsymbol{e}}_{i}^T,\boldsymbol{0}_N^T,-1,0 \right] \text{,}\;
[\tilde{\bm{J}}_{\tau}]_{i,:}=\left[\frac{\partial \mathit{h}_{\tau}(\boldsymbol{\theta})}{\partial \boldsymbol{\theta}}|_{\boldsymbol{\theta}=\tilde{\boldsymbol{\theta}}}\right]_{i,:} =\left[-\tilde{\boldsymbol{l}}_{i}^T,-\tilde{\boldsymbol{l}}_{i}^T \Delta t_i,1,\Delta t_i \right] \text{,}\;
\tilde{\boldsymbol{e}}_i=\frac{\boldsymbol{q}_i - \tilde{\boldsymbol{p}}}{\Vert \boldsymbol{q}_i - \tilde{\boldsymbol{p}}\Vert },\;
\tilde{\boldsymbol{l}}_i=\frac{\boldsymbol{q}_i - \tilde{\boldsymbol{p}}-\tilde{\boldsymbol{v}}  \Delta t_i}{\Vert \boldsymbol{q}_i - \tilde{\boldsymbol{p}}-\tilde{\boldsymbol{v}} \Delta t_i\Vert }
$.

Note that the parameter error $(\boldsymbol{\theta}-\tilde{\boldsymbol{\theta}})$ in \eqref{eq:ftheta} can be estimated in a WLS sense. Then we can refine $\tilde{\boldsymbol{\theta}}$ by a WLS step and have the final estimate $\tilde{\boldsymbol{\theta}}_{est}$ as
\begin{align} \label{eq:refinefinal}
	\tilde{\boldsymbol{\theta}}_{est}=\tilde{\boldsymbol{\theta}}+\left(\tilde{\bm{J}}^T\bm{W}\tilde{\bm{J}}\right)^{-1}\tilde{\bm{J}}^T\bm{W}\left(
	\boldsymbol{\gamma}-h(\tilde{\boldsymbol{\theta}})
	\right).
\end{align}

\section{Error Analysis} \label{locanalysis}

We first look into the estimation error of the raw estimation step (Step 2). We treat the measurement noise as perturbation on \eqref{eq:collective1} and have
$
	\left(\bm{A}+\Delta \bm{A}\right)\Delta \boldsymbol{\theta}=\Delta\boldsymbol{y}+\bm{G}\left[\Delta \lambda_1,\Delta \lambda_2\right]^T-\Delta \bm{A}\boldsymbol{\theta}
$, where the variables with ``$\Delta$'' are the error terms of the original variables caused by measurement noise. Based on \eqref{eq:H1a} and \eqref{eq:H2a}, we have $\Delta \lambda_1=2\boldsymbol{\theta}^T\bm{H}_1\Delta\boldsymbol{\theta}+\Delta\boldsymbol{\theta}^T\bm{H}_2\Delta\boldsymbol{\theta}$ and $\Delta \lambda_2=\boldsymbol{\theta}^T\bm{H}_2\Delta\boldsymbol{\theta}+\frac{1}{2}\Delta\boldsymbol{\theta}^T\bm{H}_2\Delta\boldsymbol{\theta}$. Based on the above equations and \eqref{eq:collective1}, we come to
\begin{align}
	&\left(\bm{A}+\Delta \bm{A}-\bm{G}\left[2\bm{H}_1 \boldsymbol{\theta},\bm{H}_2 \boldsymbol{\theta}\right]^T\right)\Delta \boldsymbol{\theta} \nonumber\\
	&=\bm{G}\left[\Delta \boldsymbol{\theta}^T\bm{H}_1\Delta\boldsymbol{\theta},\Delta \boldsymbol{\theta}^T\bm{H}_2\Delta\boldsymbol{\theta}/2\right]^T+\Delta \boldsymbol{y}-\Delta \bm{A}\boldsymbol{\theta}.
\end{align}

We can observe that the estimation error $\Delta \boldsymbol{\theta}$ is affected by the measurement noise contained in $\Delta \boldsymbol{y}$ and $\Delta \bm{A}$, and the true parameter $\boldsymbol{\theta}$. It is difficult to determine $\Delta \boldsymbol{\theta}$ analytically. However, through numerical simulations in the next section, we find that when the number of ANs is larger than the minimum, $\Delta \boldsymbol{\theta}$ is always small enough to ensure the optimal result in Step 3 of the CFTWLAS. 

We derive the CRLB and then compare the estimation error of the refinement step of CFTWLAS against it. The CRLB is written as
\begin{align} \label{eq:CRLB_Fisher}
	\mathsf{CRLB}
	=\left[\left(\frac{\partial h(\boldsymbol{\theta})}{\partial \boldsymbol{\theta}}\right)^T\bm{W}\frac{\partial h(\boldsymbol{\theta})}{\partial \boldsymbol{\theta}}\right]^{-1}=\bm{J}^T\bm{W}\bm{J}
	\text{,}
\end{align}
where 
$
	{\bm{J}}=[{\bm{J}}_{\rho}^T,{\bm{J}}_{\rho}^T]^T, \; [{\bm{J}}_{\rho}]_{i,:} =\left[-{\boldsymbol{e}}_{i}^T,\boldsymbol{0}_N^T,-1,0\right] \text{,}
	[{\bm{J}}_{\tau}]_{i,:}=\left[-{\boldsymbol{l}}_{i}^T,-{\boldsymbol{l}}_{i}^T \Delta t_i,1,\Delta t_i \right] \text{,}
	\tilde{\boldsymbol{e}}_i=\frac{\boldsymbol{q}_i - {\boldsymbol{p}}}{\Vert \boldsymbol{q}_i - {\boldsymbol{p}}\Vert },\;
	{\boldsymbol{l}}_i=\frac{\boldsymbol{q}_i - {\boldsymbol{p}}-{\boldsymbol{v}}  \Delta t_i}{\Vert \boldsymbol{q}_i - {\boldsymbol{p}}-{\boldsymbol{v}} \Delta t_i\Vert }.
$



We replace $(
\boldsymbol{\gamma}-h(\tilde{\boldsymbol{\theta}})$ by \eqref{eq:ftheta} into \eqref{eq:refinefinal}, and have
\begin{align} \label{eq:expand1}
	\tilde{\boldsymbol{\theta}}_{est}=&\tilde{\boldsymbol{\theta}}+\left(\tilde{\bm{J}}^T\bm{W}\tilde{\bm{J}}\right)^{-1}\tilde{\bm{J}}^T\bm{W}\left(
	\tilde{\bm{J}}\left(\boldsymbol{\theta}-\tilde{\boldsymbol{\theta}}\right)+[\boldsymbol{\varepsilon}^T,\boldsymbol{\eta}^T]^T
	\right)\nonumber\\
	=&\boldsymbol{\theta}+\left(\tilde{\bm{J}}^T\bm{W}\tilde{\bm{J}}\right)^{-1}\tilde{\bm{J}}^T\bm{W}
	[\boldsymbol{\varepsilon}^T,\boldsymbol{\eta}^T]^T,	
\end{align}
in which, the raw estimate $\tilde{\boldsymbol{\theta}}$ is eliminated.

Therefore, the covariance of the estimation error is given by
$	\mathbb{E}\left[\left(\tilde{\boldsymbol{\theta}}_{est}-{\boldsymbol{\theta}}\right)\left(\tilde{\boldsymbol{\theta}}_{est}-{\boldsymbol{\theta}}\right)^T\right]
=\left(\tilde{\bm{J}}^T\bm{W}\tilde{\bm{J}}\right)^{-1}
.$
When the number of ANs is sufficient, the geometry is appropriate, and the noise is small, $\tilde{\bm{J}}$ empirically approaches the true ${\bm{J}}$. As a result, this error covariance equals the CRLB in (\ref{eq:CRLB_Fisher}).

\section{Numerical Simulation} \label{simulation}
We first evaluate the LAS performance of the new CFTWLAS method in a 2D scene. Eight ANs are placed at the corners and the midpoints on the sides of a square area with 800 m side length. The moving UD is randomly placed inside a 500 m side-length square, which resides within and shares the same center as the AN-formed square area.

We set a varying signal-to-noise ratio (SNR), denoted by $\text{SNR}=10\log_{10}(d^2/\sigma^2)$, where $d$ is the true distance between the UD and the AN, and the unit is dB. At each SNR step, we run 10,000 Monte-Carlo simulations. During each single simulation run, the UD first transmits the request signal and then receives the response signal from AN \#$i$ after 10$i$ ms. The initial UD clock offset and drift are uniformly distributed as $b\sim \mathcal{U}(0,20)$ \textmu s and $\omega\sim \mathcal{U}(-10,10)$ parts per million (ppm), respectively. The UD velocity has a random norm drawn from $\mathcal{U}(0,50)$ m/s, and a heading angle drawn from $\mathcal{U}(0,2\pi)$.

\begin{figure}
	\centering
	\includegraphics[width=0.99\linewidth]{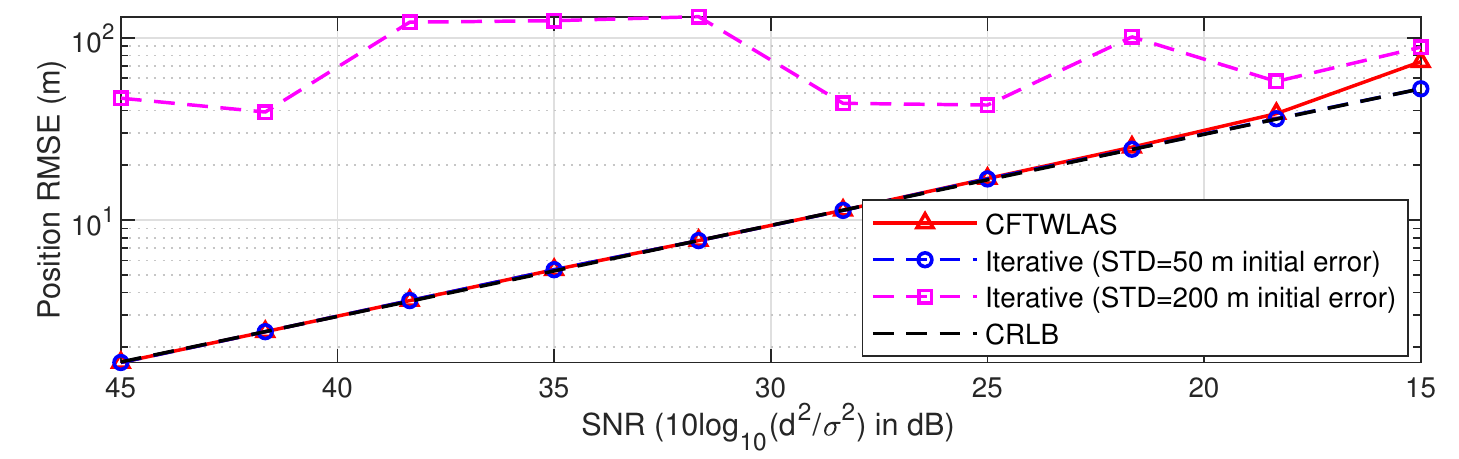}
	\vspace{-0.75cm}
	\caption{Localization error vs. SNR. The RMSE at each SNR step is the average from 10,000 simulation runs. The iterative method has large error due to inaccurate initialization (200 m STD). The position RMSE of the new CFTWLAS reaches CRLB under high SNR and deteriorates with lower SNR.
	}
	\label{fig:presultSNR}
	\vspace{-0.4cm}
\end{figure}

The localization errors from the CFTWLAS and the iterative method \cite{zhao2020optimaltwo} are shown in Fig. \ref{fig:presultSNR}. We can see that with a good initialization (50 m STD), the iterative method has optimal estimation performance. However with poorer initial position knowledge (200 m STD), which is a common case in practice, the LAS error becomes very large. For the new CFTWLAS, the results show that the position errors reach the CRLB. When the SNR falls down to about 22 dB, the errors start to deviate from the CRLB. We can change the sequential order for ANs to transmit the response signal and we have verified that the results are similar. The estimation errors for the clock offset, velocity and clock drift, which are not shown to save space, also reach CRLB.

We then investigate the performance of the new CFTWLAS with different numbers of ANs. We fix SNR=30 dB and simulate two cases - i) minimal number of 4 ANs at the square's corners, and ii) 5 ANs among which four are at the corners and one is at the middle of one side, in addition of the above 8-AN case. We compare the final localization result against the ground truth and use 3$\sqrt{\mathsf{CRLB}}$ as the threshold to identify large positioning errors. We can see from Table \ref{table_numAN} that large raw estimation errors from Step 2 will result in large errors in the final solution, especially in the case with a minimal number of ANs. However, with more ANs, large errors are less likely to appear, and the position results are reaching CRLB.

\begin{table}[!t]
	\centering
	\begin{threeparttable}
		\caption{Position Error Statistics with Different Number of ANs}
		\label{table_numAN}
		\centering
		\begin{tabular}{c  >{\raggedleft\arraybackslash}p{1.1cm}  >{\raggedleft\arraybackslash}p{0.6cm} >{\raggedleft\arraybackslash}p{0.6cm}}
			\toprule
			{ Number of ANs} & {4}&{5}&{8} \\
			\hline
			Percentage of final large error results (\%) & 16.57& 0.20 &0.00\\
			Step 2 position RMSE & 14983.23 &  32.60&20.68\\
			Final position RMSE (m) & 14952.63 &  15.22 &9.35\\
			Position CRLB (m) & 17.84 &  14.56&9.35\\
			\bottomrule
		\end{tabular}
		
		\begin{tablenotes}[para,flushleft]
			Note: SNR=30 dB. Large estimation error in Step 2 causes the large error in the final results. With more ANs, large-error results are significantly less likely to occur, and the position RMSE decreases and approaches CRLB. 
		\end{tablenotes}
	\end{threeparttable}
\vspace{-0.4cm}
\end{table}

We count the flops of the major operations  \cite{guo2021new,golub2013matrix}, and  estimate that the computation complexity is about $32N^3 +32N^2M + 104N^2+ 124NM + 148N + 130M + 697$ flops for the CFTWLAS, and about $16N^3 + 16N^2M+56N^2 + 44NM+64N + 32M + 24$ flops for one iteration of the iterative method. With the dimension $N=2$ and the number of ANs $M=8$, the CFTWLAS costs about 5,713 flops and one iteration of the iterative method takes about 1,976 flops. It shows that if the iteration count exceeds 3, the CFTWLAS has lower complexity than the iterative method.
For 10,000 simulations on PC, the actual total running time of the CFTWLAS is stable at 2.30 s. The iterative method costs 2.73 s for 3 iterations, and 4.26 s for 5 iterations. This result demonstrates that compared with the iterative method, which has increasing complexity with more iteration counts, the new CFTWLAS method has constant low complexity, and is suitable for power-constrained systems such as IoT devices.

\section{Conclusion}
We propose a new closed-form two-way TOA LAS method for a moving UD with clock drift, namely CFTWLAS. To linearize the problem, we conduct squaring and differencing on the TOA measurement equations. We then reduce the parameter estimation to finding the solution of two auxiliary variables analytically. We further refine the estimation with a WLS step. We show that with a sufficient number of ANs and a proper geometry, the estimation error of the CFTWLAS empirically reaches CRLB under the small noise condition. Numerical simulation verifies the superior performance of the new CFTWLAS. Compared with the iterative method, which requires accurate initialization to ensure correct estimation, the new CFTWLAS does not need initialization and obtains the empirically optimal result with a sufficient number of ANs as well as an appropriate geometry under small noise conditions. Furthermore, the new CFTWLAS has constant low complexity while the iterative method has growing complexity when the number of iteration increases.

\ifCLASSOPTIONcaptionsoff
    \newpage
\fi



\bibliographystyle{IEEEtran}
\bibliography{IEEEabrv,paper}
\end{document}